\documentclass[11pt]{article}
\pdfoutput=1
\usepackage{graphicx}	
\usepackage{amsmath}	
\usepackage{dcolumn}
\usepackage{bm}
\usepackage{graphics}
\usepackage{afterpage}
\usepackage{float}
\usepackage{subfigure}
\usepackage{rotating}
\usepackage{multirow}
\usepackage{tabularx}
\usepackage{booktabs}
\usepackage{multirow}
\usepackage{fancyhdr}
\usepackage[utf8]{inputenc}
\usepackage{theorem}
\usepackage{moreverb}
\usepackage{euscript}
\usepackage{psfrag}
\usepackage{slashed}
\usepackage{mathtools}
\usepackage{makecell}
\usepackage[flushleft]{threeparttable}
\usepackage{adjustbox}

\usepackage{dcolumn}
\usepackage{bm}
\usepackage[mathlines]{lineno}
\usepackage{lettrine}
\usepackage[dvipsnames]{xcolor}
\usepackage{jhep}
\usepackage{fontawesome}

\input Zallman.fd

\LettrineTextFont{\itshape}
\hypersetup{
     colorlinks   = true,
     citecolor    = PineGreen,
     urlcolor     = PineGreen,
     linkcolor    = PineGreen
}

\usepackage{listings}
\usepackage{color,xcolor}

\preprint{SLAC-PUB-17729 \\ \vspace{-8mm}\hfill LTH-1347}

\title{Dark Matter Capture in Celestial Objects: \\ Treatment Across Kinematic and Interaction Regimes}

\author[a,b]{Rebecca K. Leane,}
\emailAdd{rleane@slac.stanford.edu}
\author[c]{Juri Smirnov}
\emailAdd{juri.smirnov@liverpool.ac.uk}

\affiliation[a]{SLAC National Accelerator Laboratory, 2575 Sand Hill Rd, Menlo Park, CA 94025, USA}
\affiliation[b]{Kavli Institute for Particle Astrophysics and Cosmology, Stanford University, Stanford, CA 94035, USA}
\affiliation[c]{Department of Mathematical Sciences, University of Liverpool,
Liverpool, L69 7ZL, United Kingdom}

\date{\today}
\abstract{Signatures of dark matter in celestial objects have become of increasing interest due to their powerful detection prospects. To test any of these signatures, the fundamental quantity needed is the rate in which dark matter is captured by celestial objects. Depending on whether dark matter is light, heavy, or comparable in mass to the celestial-body scattering targets, there are different considerations when calculating the capture rate. Furthermore, if dark matter has strong or weak interactions, the physical behaviour important for capture varies. Using both analytic approximations and simulations, we demonstrate how to treat dark matter capture in a range of celestial objects for arbitrary dark matter mass and interaction strength. We release our calculation framework as a public package available in both Python and Mathematica versions, called \texttt{Asteria}~\href{https://doi.org/10.5281/zenodo.8150110}{\faExternalLink}.
}

\usepackage{natbib}
\usepackage{graphicx}

\begin{document}
\maketitle

\newpage
\section{Introduction}

\lettrine{C}{apture of dark matter} (DM) in celestial objects gives rise to wide-ranging and exciting opportunities for DM discovery. Assuming interactions with the Standard Model (SM) particles, DM may scatter with the celestial-body matter, lose energy, and become gravitationally captured. This can lead to the build up of a DM population inside the celestial object, with readily detectable signals. For annihilating DM, these include heating signals, neutrino signals, gamma-ray signals, or electron signals. For non-annihilating DM, very large amounts of DM can accumulate over time due to a lack of depletion through annihilation, leading to interesting observables such as eventual destruction via black holes from over-accumulation, or changes to stellar evolution. The range of objects considered in the past includes the Earth and the Sun~\cite{Batell:2009zp,Pospelov:2007mp,Pospelov:2008jd,Rothstein:2009pm,Chen:2009ab,Schuster:2009au,Schuster:2009fc,Bell_2011,Feng:2015hja,Kouvaris:2010,Feng:2016ijc,Allahverdi:2016fvl,Leane:2017vag,Arina:2017sng,Albert:2018jwh, Albert:2018vcq,Nisa:2019mpb,Niblaeus:2019gjk,Cuoco:2019mlb,Serini:2020yhb,Acevedo:2020gro,Mazziotta:2020foa,Bell:2021pyy}, Jupiter~\cite{Batell:2009zp,Leane:2021tjj,Li:2022wix,French:2022ccb,Ray:2023auh}, Brown Dwarfs~\cite{Leane:2020wob,Leane:2021ihh}, Uranus~\cite{Mitra:2004fh}, Exoplanets~\cite{Leane:2020wob}, White Dwarfs and Neutron Stars~\cite{Goldman:1989nd,
Gould:1989gw,
Kouvaris:2007ay,
Bertone:2007ae,
deLavallaz:2010wp,
Kouvaris:2010vv,
McDermott:2011jp,
Kouvaris:2011fi,
Guver:2012ba,
Bramante:2013hn,
Bell:2013xk,
Bramante:2013nma,
Bertoni:2013bsa,
Kouvaris:2010jy,
McCullough:2010ai,
Perez-Garcia:2014dra,
Bramante:2015cua,
Graham:2015apa,
Cermeno:2016olb,
Graham:2018efk,
Acevedo:2019gre,
Janish:2019nkk,
Krall:2017xij,
McKeen:2018xwc,
Baryakhtar:2017dbj,
Raj:2017wrv,
Bell:2018pkk,
Chen:2018ohx,
Dasgupta:2019juq,
Hamaguchi:2019oev,
Camargo:2019wou,
Bell:2019pyc,
Acevedo:2019agu,
Joglekar:2019vzy,
Joglekar:2020liw,
Bell:2020jou,
Dasgupta:2020dik,
Garani:2020wge,
Leane:2021ihh,Collier:2022cpr}, and other stars~\cite{Freese:2008hb, Taoso:2008kw, Ilie:2020iup, Ilie:2020nzp, Lopes:2021jcy,Ellis:2021ztw}.

To test any of these signatures, the fundamental quantity needed is the DM capture rate. Formalisms to calculate the capture rate of DM go back a very long time~\cite{1985ApJ296679P, Gould:1987ir}, and have recently been improved predominantly in the context of heavy DM in high-escape velocity objects, such as neutron stars or white dwarfs~\cite{Bramante:2017xlb,Bell:2019pyc,Bell:2020jou,Bell:2020obw,Bell:2021fye}. However, in recent years, interest has grown in signatures of objects with low escape velocities, such as the Earth or Jupiter, due to highly detectable signals~\cite{Leane:2021tjj,Li:2022wix,French:2022ccb,Ray:2023auh,Neufeld:2018slx,Pospelov:2020ktu, Pospelov:2019vuf, Rajendran:2020tmw,Xu:2021lmg,Budker:2021quh, McKeen:2022poo,Billard:2022cqd, Leane:2022hkk,Das:2022srn,McKeen:2023ztq}. Interest has also recently grown in the prospects for light DM discovery, due to strengthening direct detection bounds for GeV-scale DM~\cite{LUX-ZEPLIN:2022qhg,Hochberg:2022apz}. This is especially pertinent as it was recently shown that even sub-MeV DM masses can be retained in a wide range of celestial objects, depending on the DM model~\cite{Acevedo:2023owd}. Capture of light DM with strong interactions in the Earth was analytically approximated in Ref.~\cite{Neufeld:2018slx}. However, any framework considering comparable DM and SM target masses, or light DM with cross sections near the strong-weak interaction boundary, have not been considered. There are also a number of other regimes which have not yet been explored in full generality, as well as some inconsistencies in the existing literature.

\begin{figure}[t!]
\centering
\includegraphics[width=0.5\columnwidth]{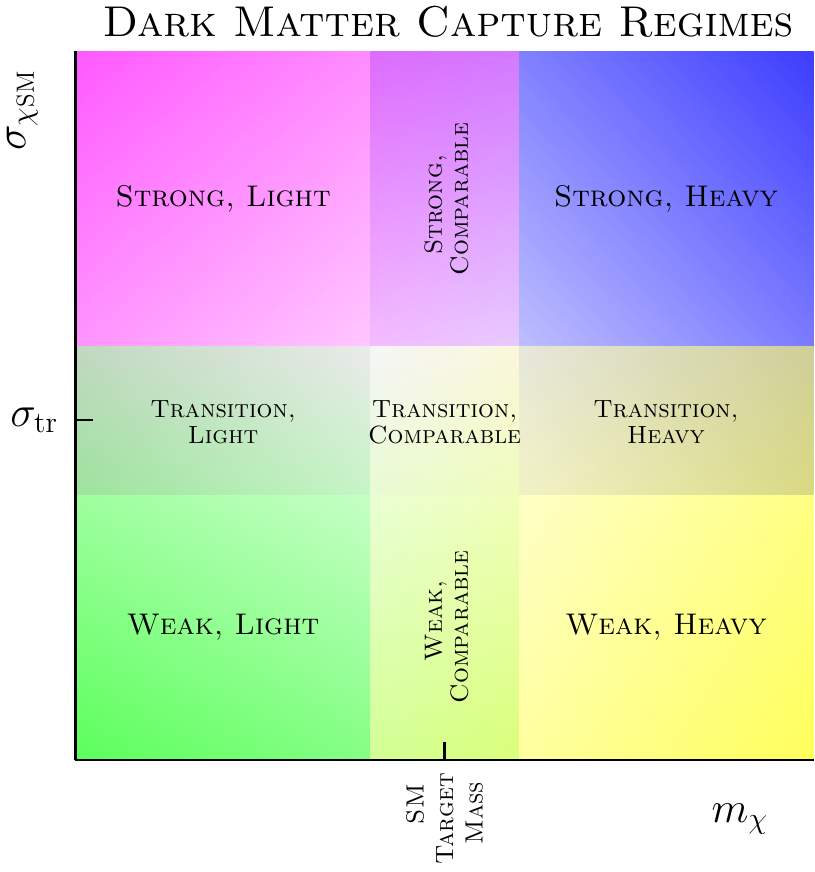}
    \caption{Schematic of the DM capture regimes of interest, as a function of DM-SM scattering cross section $\sigma_{{\chi}{\rm SM}}$ and DM mass $m_\chi$. Each regime can have unique features requiring different capture treatments. The transition cross section $\sigma_{\rm tr}$ is the cross section corresponding to a mean free path of the size of the celestial object, and therefore the transition between the single and multi-scatter interaction regime. The DM mass relative to the celestial body SM target mass also affects the capture kinematics and features.}
    \label{fig:capturegridfull}
\end{figure}

The goal of this work is to establish capture rate calculations which are valid for a wide range of celestial objects, in arbitrary DM mass regimes, and arbitrary DM-SM scattering cross sections, covering all the regions in our Fig.~\ref{fig:capturegridfull}, and focusing on DM-nucleon interactions. To do this, we will present both analytic and simulated results, and present some conditions on how to treat DM capture. We find our improved treatment can lead to results that are very different to other frameworks. Our framework is released in a public package available in both Python and Mathematica versions, called \texttt{Asteria}~\cite{asteria}, named after the Greek goddess of stars and planets.

Our paper is organized as follows. In Section~\ref{sec:overview}, we give an overview as well as some improvements in the foundations of the main existing multiscatter formalism, and detail some treatments for the capture summation in Section~\ref{sec:sum}. In Section~\ref{sec:weak} we briefly discuss the weak interaction regime. In Section~\ref{sec:diffuse}, we discuss the diffusive DM regime, corresponding to DM with light or comparable to SM target masses, with large interaction cross sections. This includes new simulation results. In Section~\ref{sec:transition}, we discuss the intermediate ``transition" interaction regime, where the DM-SM interaction rate switches between the single and multi-scatter treatments. In Section~\ref{sec:ballistic}, we discuss treatment of the heavy ballistic DM regime, and in Section~\ref{sec:objects} we apply our capture formalism to the Earth, the Sun, Jupiter, and a Brown Dwarf, and describe the physical features observed throughout each capture regime. We summarize our findings in Section~\ref{sec:conclusion}.

\section{Overview of Capture Formalism}
\label{sec:overview}

\subsection{Previous Multiscatter Formalism Assumptions}

 DM capture can occur via single or multiple scatters in the celestial body, depending on the kinematic regime. A treatment covering mulitscatter of DM was first studied in Ref.~\cite{Gould:1987ir}, and a currently widely used framework is from Ref.~\cite{Bramante:2017}, though also see Ref.~\cite{Ilie:2020}. Further improvements were also shown for ultra-dense objects in Refs.~\cite{Bell:2018pkk,
Bell:2019pyc,Bell:2020jou,Bell:2021fye}. However, as the framework of Ref.~\cite{Bramante:2017} was created for heavy DM in neutron stars, it is not generically applicable to all celestial objects in all DM mass regimes. In particular, the assumptions of Ref.~\cite{Bramante:2017} are that:
 \begin{itemize}
  \item The celestial-body is a neutron star, such that its escape velocity $v_{\rm esc}$ is much greater than the mean incoming halo DM velocity $v_\chi$, i.e. $v_{\rm esc}\gg v_\chi$,
  \item The DM is heavy compared to the SM targets in the celestial body. This is because the assumption is that DM travels in a straight line through the object, corresponding to ballistic forwards motion, which does not apply to light DM.
 \end{itemize}

 For strongly interacting light DM in the Earth, an analytic modification was proposed in Ref.~\cite{Neufeld:2018slx}. This can be used for DM much lighter than the SM target matter, but does not apply in the weak interaction transition nor weak interaction regime. No framework exists for comparable DM and SM masses outside the weak-interaction regime. We will address these cases shortly, but first we now discuss the foundation for the capture calculation, which will serve as a backbone for a few of our improved treatments.

\subsection{Foundation for the Capture Calculation}

As we aim to detail capture for a range of objects, DM masses, and DM-SM interaction regimes, we start by giving an overview of the minimal ingredients for capture, under the assumption of forward linear DM motion; we will address the modifications required for specific regimes in the upcoming sections.

The probability of capture after $N$ scatters, assuming isotropic scattering (i.e. that the differential cross section is independent of the scattering angle), is given by 
\begin{align}
\label{eq:gNfull}
    g_N(u) & = \int_0^1 dz_1 \, ... \int_0^1 dz_N \, \Theta \left(1 - \sqrt{1 + w^2} \, \prod_{i=1}^{i=N} \sqrt{ 1 - z_i \beta } \right)\,,
\end{align}
with $z_i$ being the angular scattering variables, $w = u/v_{\rm esc}$, where $u$ is the velocity of the incoming DM particle, and
\begin{equation}
    \beta = \frac{4 m_{\rm SM}\, m_{\chi}}{(m_{\rm SM} + m_{\chi})^2},
\end{equation}
where $m_{\rm SM}$ is the SM target mass, and $m_\chi$ is the DM mass.
We define $A =\prod_{i=1}^{i=N-1} \sqrt{ 1 - z_i \beta }$, and use it rewrite the $N$th integral. This integral can be evaluated analytically, however, to do this the required condition is that
\begin{align}
\label{eq:capconditions}
& A \,\sqrt{u^2 + v_{\rm esc}^2} (1 - \beta)^{1/2} < v_{\rm esc} \, , \\ \nonumber
& \text{and } A \, \sqrt{u^2 + v_{\rm esc}^2} > v_{\rm esc}.
\end{align}
The condition in the first line of Eq. (\ref{eq:capconditions}) implies that the maximal energy transfer possible in the $N$th scatter with $z = 1$ brings the DM velocity below the escape velocity such that DM is captured. The condition in the second line implies that the DM particle is still above the escape velocity after the previous $(N-1)$th scatter. Thus, as pointed out in Ref.~\cite{Leane:2022hkk}, the evaluated integral gives the probability for getting captured after exactly $N$ collisions, and reads
\begin{align}
      \int_0^1 dz_N \, \theta \left(1 - \sqrt{1 + w^2} \, A \, \sqrt{ 1 - z_N \beta } \right) = 1 - \frac{1}{\beta} + \frac{1}{\beta \, (1 + w^2) A^2}\,.
\end{align}
In Ref.~\cite{Dasgupta:2019juq}, this expression has been evaluated, and after further $N-1$ integrations is 
\begin{align}
    \left. g_N(u)\right|_{\rm Exactly\ N} & = \int_0^1 dz_1 \, ... \int_0^1 dz_{N-1} \left( 1 - \frac{1}{\beta}  + \frac{1}{ \left( 1+ w^2 \right) \prod_{i=1}^{i=N-1} \left( 1 - z_i \beta \right)  \beta} \right) \\
    &= 1 - \frac{1}{\beta}  + \frac{1}{ \beta^N \left( 1+ w^2 \right) } \log{ \left[\frac{1}{1 - \beta}\right]}^{N-1}\,.
    \label{eq:gnexact}
\end{align}
The capture rate after $N$ or less scatterings is then obtained by summation over $N$ and integration over the DM velocity distribution~\cite{Leane:2022hkk}
\begin{align}
    C_N = \pi R^2  p_N(\tau) \, \sum_{i=1}^{ N} \int_0^\infty f(u)  u  \left( 1 + w^{-2} \right) \,\left.g_i(u)\right|_{(\rm Exactly\ i)}\, du  \,.
    \label{eq:cnhm}
\end{align}
To get the total capture rate one needs to sum the coefficients in Eq.~(\ref{eq:cnhm}) up to a maximal number of scatterings $N_{\rm max}$,
\begin{align}
\label{eq:CNtotal}
   C_{\rm total} = \sum_{i=1}^{N_{\rm max}} C_i\,.
\end{align}
$N_{\rm max}$ is a truncation condition used to allow numerical evaluation, which we discuss in the next section. Note that while Eq.~(\ref{eq:gnexact}) was evaluated in Ref.~\cite{Dasgupta:2019juq}, that reference misses the summation in Eq.~(\ref{eq:cnhm}) and therefore produces unphysical behaviour; see Appendix B of Ref.~\cite{Leane:2022hkk} for details.

The expression in Eq.~(\ref{eq:cnhm}) is greatly simplified in the large interaction regime, where the DM undergoes many scatterings.  The nested integral in Eq.~(\ref{eq:gNfull}) can be evaluated using the assumption that the scattering variable is on average $\langle z_i \rangle \approx 1/2$, which yields an approximate result for $g_N(u) \approx g_N(u)^{\rm avg}$, as discussed in Ref.~\cite{Bramante:2017}. The expression $g_N(u)^{\rm avg}$ corresponds to the probability for the DM particle to drop below escape velocity at any of the $N$ scatterings, as no additional kinematic conditions are needed to evaluate the integral. Therefore, no additional summation is required and integrating $g_N(u)^{\rm avg}$ over the DM velocity distribution 
\begin{align}
    C_N \approx \pi R^2 p_N(\tau) \int_0^\infty f(u)  u  \left( 1 + w^{-2} \right) g_N(u)^{\rm avg}\, du  \,,
    \label{eq:cnbram}
\end{align}
we obtain a compact analytic expression for the capture coefficient 
\begin{equation}
\label{eq:multis}
C_N = C_{\rm geo} \, p_N(\tau) \, \left( 1 - \frac{\exp \left[ -\frac{3}{2}\frac{v_{\rm esc}^2}{v_\chi^2} \left( \alpha_\mu^{-N} - 1 \right) \right] \left( 1 + \frac{3}{2} \frac{v_{\rm esc}^2}{v_\chi^2} \alpha_\mu^{-N} \right)}{1 + \frac{3}{2}\frac{v_{\rm esc}^2}{v_\chi^2}} \right)\, ,
\end{equation}
where the geometric capture rate $C_{\rm geo}$ is the total rate of DM particles passing through the celestial body 
\begin{equation}
    C_{\rm geo} = \frac{\pi R^2 \rho_\chi v_\chi }{m_\chi} \sqrt{\frac{8}{3 \pi}} \left( 1 + \frac{3}{2} \frac{v_{\rm esc}^2}{v_\chi^2}  \right) \, ,
\end{equation}
 where $\rho_\chi$ is the DM density in the surrounding environment, as found in Ref~\cite{Bramante:2017}. The energy fraction remaining after each scatter is given by
\begin{equation}
   \alpha_\mu = 1- \frac{4 \langle z \rangle \mu}{(1+\mu)^2},
\end{equation}
with $\mu = m_\chi/m_{\rm SM}$.
 
 We use the full expression for arbitrary scattering angle in the weak interaction limit, as the averaged scattering angle is only a good approximation for multi-scatter interactions. In the case of large interactions and multi-scatter, the $\langle z_i \rangle \approx 1/2$ approximation speeds up and the calculation without a relevant loss of accuracy (see Fig.~5 in App. B of Ref.~\cite{Leane:2022hkk}), and so we use Eq.~(\ref{eq:multis}) for the capture coefficient, to get an expression for the capture rate after $N$ or less scatters, which is then summed over per Eq.~(\ref{eq:CNtotal}).

In the above form of Eq.~(\ref{eq:multis}) it is easy to see that the capture rate for all $C_N$ drops to zero if $\mu \rightarrow 0$ or $\mu \rightarrow \infty$, as $\alpha_\mu \rightarrow 1$, i.e. the remaining energy fraction after each scattering approaches one. This shows that the energy loss efficiency is poor both for increasingly light or heavy DM mass compared to the target mass, such that more scatters are increasingly required in these limits. Note that we neglect the relative motion of the celestial object to the DM halo, which is a small correction. We also neglect thermal motion of the targets, which is justified for interactions with nuclei in most celestial objects given sufficient DM mass.

The probability of a single DM particle undergoing $N$ scatters is~\cite{Bramante:2017}
\begin{equation}
    p_N(\tau) = 2 \int_0^1 dy \frac{y e^{-y \tau} (y \tau)^N}{N!},
    \label{eq:pn}
\end{equation}
where $y$ is an angular impact variable and $\tau$ is the optical depth,
\begin{align}
    \tau  = \sum_i \,\frac{3}{2} \frac{\sigma_{\chi_{A_i}}}{\sigma_{\text{tr}_{A_i}}} =\sum_i \tau_{A_i},
\end{align}
where $\sigma_{\text{tr}_A} = \pi R^2/N_{A}$ is the transition cross section for DM capture, with $N_A = f_A M_{\rm obj}/m_A$ being the number of atoms with atomic mass number $A$ in the object, with $f_A$ being the mass fraction of the element. The transition cross section is when the mean free path of the DM is the diameter of the celestial object, and therefore is at the transition of the single and multiscatter regimes. Note that $\sigma_{\rm tr}$ has been called the ``saturation cross section" $\sigma_{\rm sat}$ in previous works, and we advocate for renaming it as calling it the saturation cross section is not physically accurate for all objects; we discuss this in more detail in the upcoming Sec.~\ref{sec:sat}.

For spin-independent scattering rates in celestial objects, we convert the nucleon to nucleus cross sections using the Born approximation,
\begin{align}
   \sigma_{\chi \rm A} = A^2 \left(\frac{\mu_R(A)}{\mu_R(N)}\right)^2{\sigma_{\chi N} },
   \label{eq:born}
\end{align}
where $A$ is the atomic mass number of the relevant nucleus, and $\mu_R(A \text{ or } N)$ are the reduced masses of the DM with the nucleus or nucleon. Note that the Born approximation breaks down for large cross sections, and in this regime particle DM models should be used directly~\cite{Digman:2019wdm,Xu:2020qjk}. For practical purposes an approximation procedure allows preservation of unitarity of the interaction by assuming $\sigma_{\chi A}^{\rm tot} = \min \left(\sigma_{\chi A} , 4 \pi r_N^2 \right) $, where $r_N \approx (1.2\, \text{fm})\,A^{1/3}$~\cite{Digman:2019wdm}. However, if DM is an extended object with $r_\chi > r_N$, larger cross sections are possible. To remain model agnostic we do not limit the cross section size, but note that depending on the model, the treatment of large cross sections ($\sigma_{\chi N} \gtrsim 10^{-31}$~\cite{Digman:2019wdm}) can substantially deviate from our setup here. In our package \texttt{Asteria}~\cite{asteria} there is the option to include the $A$ scaling, however we emphasize that this will not be valid for large cross sections. We therefore also provide the possibility for the input of the total $\sigma_{\chi A}$ cross section, such that no scaling assumption with atomic mass number is enforced internally.

\subsection{Multi-Element Treatment}

Treating capture for multi-component celestial objects can quickly become cumbersome, due to the large number of scattering combinatorics. We make a few simplifying assumptions, without sacrificing much accuracy. Firstly, we assume a homogeneous distribution of the chemical elements throughout the celestial body. We discuss our additional assumptions below depending on the regime.

\subsubsection{Multi-Scatter Case}
As discussed in Ref.~\cite{Ilie:2021iyh} in the context of a two component model in the optically thick regime, the scattering combinatorics can start to play a role. We find that a sufficient treatment is to switch at cross sections above the transition cross section i.e.  $\tau > 3/2$, to the effective treatment using an average mass, dominated by the element that is most likely to be encountered by DM
\begin{equation}
     m_{\rm SM} = \frac{1}{\tau} \sum_i \tau_{A_i} \, m_i\,.
\end{equation}
This treatment greatly accelerates the computation compared to evaluating all the combinatorics, and produces results in an agreement within a few percent with the full approach in Ref.~\cite{Ilie:2021iyh}.

\subsubsection{Single-Scatter Case}

In the case that $\tau < 3/2$ the scattering is dominated by the first, or first few, scattering events. In this regime it is sufficient to calculate the capture rates per element  $C_{A_i}$, which are dependent on the $\tau_{A_i}$ for each element independently, and sum them to obtain a total capture rate
\begin{align}
    C_{\rm all \, elements} =  \sum_i C_{A_i} \,.
    \label{eq:multiweak}
\end{align}
This treatment agrees within percent level accuracy with the full calculation that takes into account combinatorics, see Ref.~\cite{Ilie:2021iyh}. 

\section{How to Treat the Capture Summation}
\label{sec:sum}

The multi-scatter sum has formally an infinite number of elements, thus in order to evaluate it numerically we use a truncation condition, and several simplifications to accelerate the calculation in certain regimes. We will first formulate and justify the truncation condition, defining the maximum number of scatters, and discuss regime specific simplifications in the following dedicated sections.

In order to formulate the truncation condition it is crucial to understand the behaviour of the probability of DM to scatter $N$ times in the object. While the expression for the capture rate at large cross section in Eq.~(\ref{eq:CNtotal}) is correct, it can be non-trivial to numerically compute, particularly due to the $p_N(\tau)$ as written in Eq.~(\ref{eq:pn}), which can require evaluating large $N!$. We now discuss the behavior of the scattering probability at large optical depth, that will allow us to formulate a truncation condition for the capture rate sum.

To improve numerical stability in the strong interaction regime, it is advantageous to write Eq.~(\ref{eq:pn}) as~\cite{Ilie:2020}
\begin{equation}
    p_N(\tau) = \frac{2(N+1)}{\tau^2} \left( 1 - \frac{\Gamma[ N+2 , \tau]}{\Gamma[N+2]}\right),
    \label{eq:pnfixed}
\end{equation}
where $\Gamma[a]$ is Gamma function, and $\Gamma[a,b]$ is the upper incomplete Gamma function; gamma functions are used as they are known as the generalized factorial, as $\Gamma[N+1]=N!$ for natural numbers $N$.

\begin{figure}[t!]
\centering
\includegraphics[width=0.6\columnwidth]{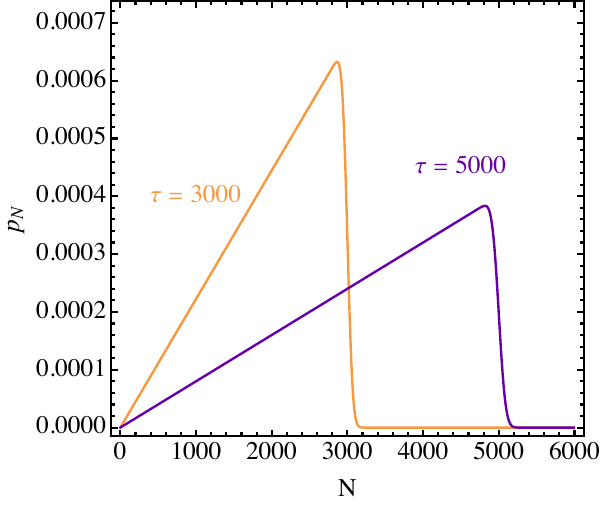}
    \caption{Probability $p_N$ of exactly $N$ scatters for different choices of the optical depth $\tau$.}
    \label{fig:pN}
\end{figure}

Figure~\ref{fig:pN} shows the scatter probability as a function of the number of scatters for different choices of optical depth $\tau$, 3000 and 5000. We see that the probability to scatter $N$ times peaks around $N \sim \tau$, which is true for all $\tau \gg 3/2$. This means that the probability to scatter many more times than the optical depth is negligible, in the strongly interacting regime. We make this point more explicit by considering an expansion of Eq.~(\ref{eq:pnfixed}) in two regimes

\begin{align}
\label{eq:pscatterexp}
    p_{N}(\tau) \approx  \frac{2 (N+1)}{\tau^2} \times 
    \begin{cases}
         \left(1 - \dfrac{e^{-\tau} \tau^{1 + N}}{\Gamma[N+2]}\right)  & (N < \tau) ,\\ \\
        \left(\dfrac{e^{N-\tau} N^{-5/2-N} \tau^{N+2}}{\sqrt{2 \pi}}\right) &  (N > \tau)\,.
    \end{cases}
\end{align}
In the limit that $\tau\gg3/2$, the right bracket for the $N<\tau$ case approaches one, while the bracket for $N>\tau$ approaches zero, i.e.
\begin{equation}
\label{eq:pNstep}
    p_{N}(\tau) \approx  \frac{2 (N+1)}{\tau^2}\Theta(\tau-N),
\end{equation}
in agreement with Ref.~\cite{Ilie:2020}. This shows that for large cross sections only $N < \tau$ meaningfully contributes to the total capture rate. If instead we are in the weakly interacting regime, i.e. if $\tau$ is smaller than about $3/2$, the probability function  $p_N(\tau)$ peaks at $N = 1$ (meaning one scatter) and rapidly decays with growing $N$.

In the package that we provide along with this paper, the numerical accuracy goal for $p_N$ is at the percent level, and so we find that it is computationally sufficient to truncate the summation in Eq.~(\ref{eq:CNtotal}) at
\begin{equation}
    N_{\rm max} = \max \lbrace  10, \lfloor e \tau \rfloor  \rbrace\,.
\end{equation}
This is an arbitrary choice, which is justified because at $N \sim e \tau$, Eq.~(\ref{eq:pscatterexp}) for $\tau < N$  leads to a scaling for $p_N(\tau) \sim e^{-\tau}/\tau^{5/2}$, which is negligible for large $\tau$ values. For smaller optical depth, the capture is controlled by the first few scatters, since the at small $\tau < 3/2$ Eq.~(\ref{eq:pnfixed}) leads to a rapidly falling scattering probability $p_N(\tau) \sim \tau^{N}/N!$. We make thus a conservative choice if we sum to $N_{\rm max} =10$ in this case.

There is one more caveat concerning the optical depth, which is that there are a finite number of SM targets to scatter with. This is generally only relevant in the ultra-heavy DM limit; we discuss it further in Sec.~\ref{sec:ballistic}.
 
By implementing the above conditions we find that the numerical evaluation of Eq.~(\ref{eq:CNtotal}) is greatly accelerated, and is within a sub-percent level accuracy of the fully converged general result.

We now discuss regime-specific calculations for DM capture rates.
 
\section{Weak Interaction Regime}
\label{sec:weak}

\begin{figure}[H]
\centering
\includegraphics[width=0.5\columnwidth]{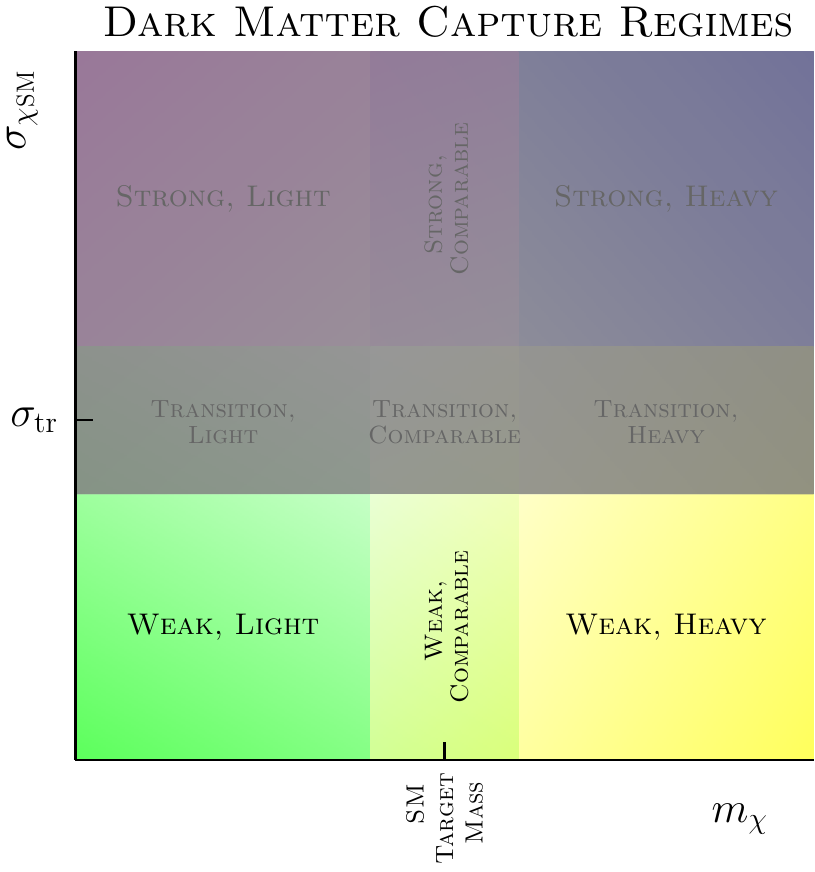}
    \caption{The weak regime discussed in this section.}
    \label{fig:capturegridlightstrong}
\end{figure}

The optically thin regime, also known as the weak interaction regime, has been the most commonly studied. The original results of Gould~\cite{Gould:1987ir} (see also Spergel and Press~\cite{1985ApJ296679P}) hold in this regime, but are also approximately reproduced in the limiting case of weak cross sections in the multiscatter calculation in Eq.~(\ref{eq:CNtotal}). Here, there is no need to deviate from the simple non-linear forward scattering picture described in Sec.~\ref{sec:overview}, as the DM may change direction, but it only does so roughly once (as far as capture is concerned), making the direction it turns largely irrelevant. 

\subsection{Numerical Simplifications}

When taking the single scatter limit of Ref.~\cite{Bramante:2017}, one point that potentially requires some care is the evaluation of the scattering probability. It can be numerically advantageous to use the leading order expansion of Eq.~(\ref{eq:pscatterexp}) for $\tau \ll 1$,
\begin{align}
    p_N(\tau) \approx \frac{2\, \tau^N (N +1 )}{(N+2) \Gamma[N+2]}\,.
\end{align}

In addition, when the mass mismatch between the DM and the SM target is very large i.e. $\mu \ll 1$ or $\mu \gg 1$ the capture rate coefficients are $C_N/G_{\rm geo} \ll 1$ and can become numerically unstable in the weak regime. In this case an approximate expression can be used to alleviate the problem,
\begin{align}
\label{eq:CNlow}
   C_N \approx 9 \, C_{\rm geo}\,p_N(\tau) N\,   \frac{v_{\rm esc}^4}{v_\chi^4 } \frac{\mu}{(1+ \mu)^2}\, \left(2 + 3 \frac{v_{\rm esc}^2}{v_\chi^2} \right)^{-1} \,.
\end{align}
This expression is valid for $\mu \ll 1$ and $\mu \gg 1$, which is the case for inefficient energy transfer, given a large mass mismatch.  When in the small optical depth regime, we replace $C_N$ by the approximate expression in Eq.~(\ref{eq:CNlow})  when $\mu/(1 + \mu^2) \times v_{\rm esc}^2/v_\chi^2 < 10^{-5}$ which improves the numerical stability and allows us to evaluate even highly suppressed capture rates.

\section{Diffusive Dark Matter Regime}
\label{sec:diffuse}

\begin{figure}[H]
\centering
\includegraphics[width=0.5\columnwidth]{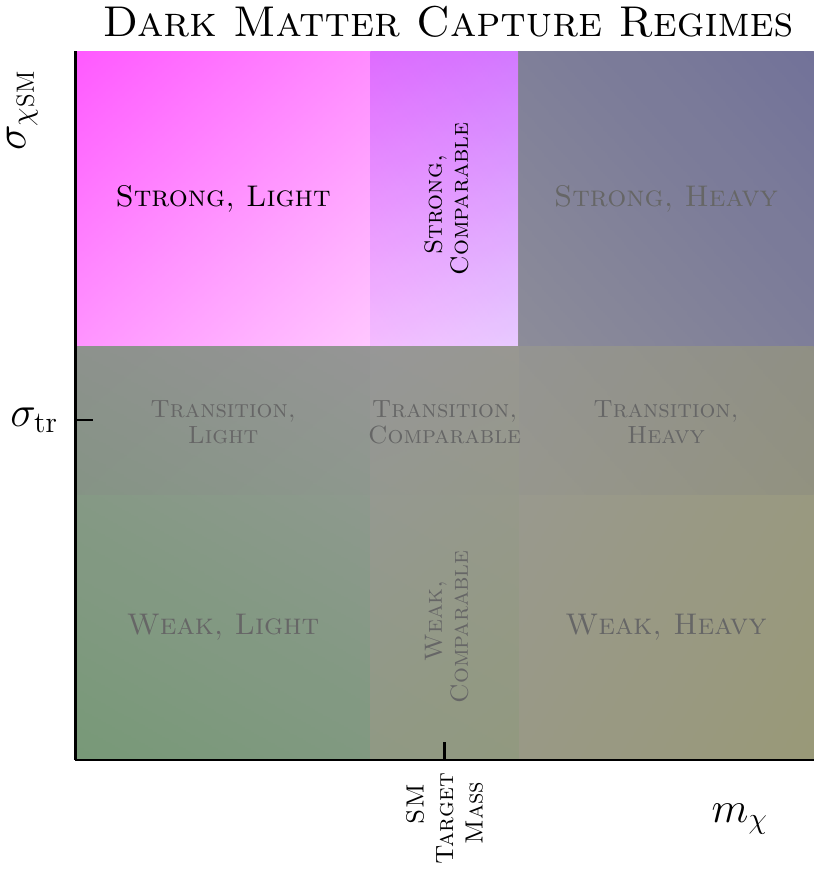}
    \caption{The strong light and strong comparable regime discussed in this section.}
    \label{fig:capturegridstronglightcomp}
\end{figure}

As discussed above, the capture ingredients in Sec.~\ref{sec:overview} assume that all DM trajectories are a straight forward line through the object. This holds for DM much heavier than the SM target, or weakly interacting DM that is much lighter than the SM target. 

On the other hand, when DM has strong interactions and is light or comparable in mass to the SM target, it has diffusive motion while being captured, and can be more easily have its trajectory altered backwards or sideways~\cite{Gould:1989hm,Zaharijas:2004jv,Neufeld:2018slx}. In the multi-scatter regime (i.e., the strong interaction regime), these non-forwards trajectories need to be tracked. To include this effect in Eq.~(\ref{eq:multis}), we will modify the expression with an additional factor $f_{\rm cap}$, which takes into account reflection of light DM out of the celestial object. This is required as Eq.~(\ref{eq:multis}) only applies for heavy DM in celestial objects with high escape velocities, and we want to extend the setup to include light DM masses in objects with low escape velocities. In addition, as we will show, this diffusive motion factor can also actually be important for high-escape velocity objects such as the Sun when DM becomes very light, as the DM is sped up to the escape velocity and can still end up reflected back out of the object.

\subsection{Required Number of Scatters for Maximal Dark Matter Capture}

To calculate the amount of reflection, we will need an estimate of the expected number of scatters before capture. 
The number of scatters at which the majority of DM particles are moving at an average velocity was derived in Ref.~\cite{Mack:2007xj}, however, the only the halo velocity was considered for the incoming DM velocity. We extend the formula to take into account the effect of the gravitational infall i.e. $v_{\rm in}  = \sqrt{v_\chi^2 + v_{\rm esc}^2}$, which is important for heavy objects. We therefore find the number of scatterings that is required for the bulk of the DM particles to be captured,
\begin{align}
  N_{\rm req}(\mu, v_\chi,v_{\rm esc})  =  \frac{ \log\left( v_{\rm esc}^2/(v_\chi^2+v_{\rm esc}^2) \right)}{\log \left( \alpha_\mu \right)} \,,
  \label{eq:nreq}
\end{align}
which depends on the escape velocity and average DM velocity, and the DM mass and the average target mass, and where again the average of a scattering variable $\langle z\rangle \sim 1/2$ has been taken assuming isotropic scattering. From Eq.~(\ref{eq:nreq}), it is clear that if the DM mass is significantly larger or smaller than the SM target mass, multiple scatterings have to be taken into account in order to correctly compute the DM capture rate.

\subsection{Reflection Coefficient in the Low Dark Matter Mass Limit}

The simplest case to consider including reflection is the limit that the DM is much lighter than the SM target. In this case, the reflection factor is given by
\begin{align}
\label{eq:fcap}
   f_{\rm cap}^{\rm light}\approx \frac{2}{ \sqrt{\pi\, N_{\rm req}}} =\left[ \frac{4}{\pi}  \frac{\log \left( \alpha_\mu \right) }{\log\left( v_{\rm esc}^2/(v_\chi^2+v_{\rm esc}^2) \right)}
 \right]^{1/2},
\end{align}
where $N_{\rm req}$ is the total number of scatters required for capture (distinct from $N$, which is instead the total number of scatters the DM actually undertakes). A similar expression was pointed out in Ref.~\cite{Neufeld:2018slx} for the Earth, though there the DM speed up from the escape velocity was not included, which makes a non-negligible difference for higher escape velocity objects. In any case, as the DM mass approaches the SM target mass, this approximation becomes less reliable, and so far there was no analytic expression to be used. We now investigate more general behavior of the reflection coefficient $f_{\rm cap}$.

\subsection{New Simulations and Analytic Fits}

\subsubsection{Simulation Setup}

We perform simple simulations to investigate the generic behaviour of the reflection coefficient. Our simulation setup is similar to that performed in Ref.~\cite{Banks:2021sba} for the DM distribution within the Sun, although we will apply our simulation to a different problem which is the fraction of reflected DM particles, rather than the internal DM distribution. 

We perform a 3D Monte Carlo Simulation assuming the DM particles enter the celestial object with a velocity of $v_{\rm in} = \sqrt{v_{\rm esc}^2 + v_\chi^2}$, and at an impact vector randomly drawn from a distribution uniform in angles. In order to determine the next interaction point, including the first point after entry, we draw the realized free path from an exponential distribution proportional to $e^{-\tau}$, where $\tau$ is the optical depth. We perform the DM particle scattering in the center of mass frame, which conserves the magnitude of the velocity vector. We assume isotropic scattering, thus drawing the new direction angles from a flat distribution, and boost to the lab frame of the object. The simulation is run until the DM particle either leaves the object, or its velocity drops below $v_{\rm esc}$, in which case it is considered captured. We sample over $10^4$ particles per mass point, as this number of particles exhibits well converged behavior. This gives us our data for our analytic fit to the reflection coefficient. The simulations are valid in the optically thick regime, as the mean free path is limited to near the surface of the object. We therefore simplify the setup by only taking the $v_{\rm esc}$ value at the surface.

\begin{figure*}[t!]
    \centering
     \includegraphics[width=\columnwidth]{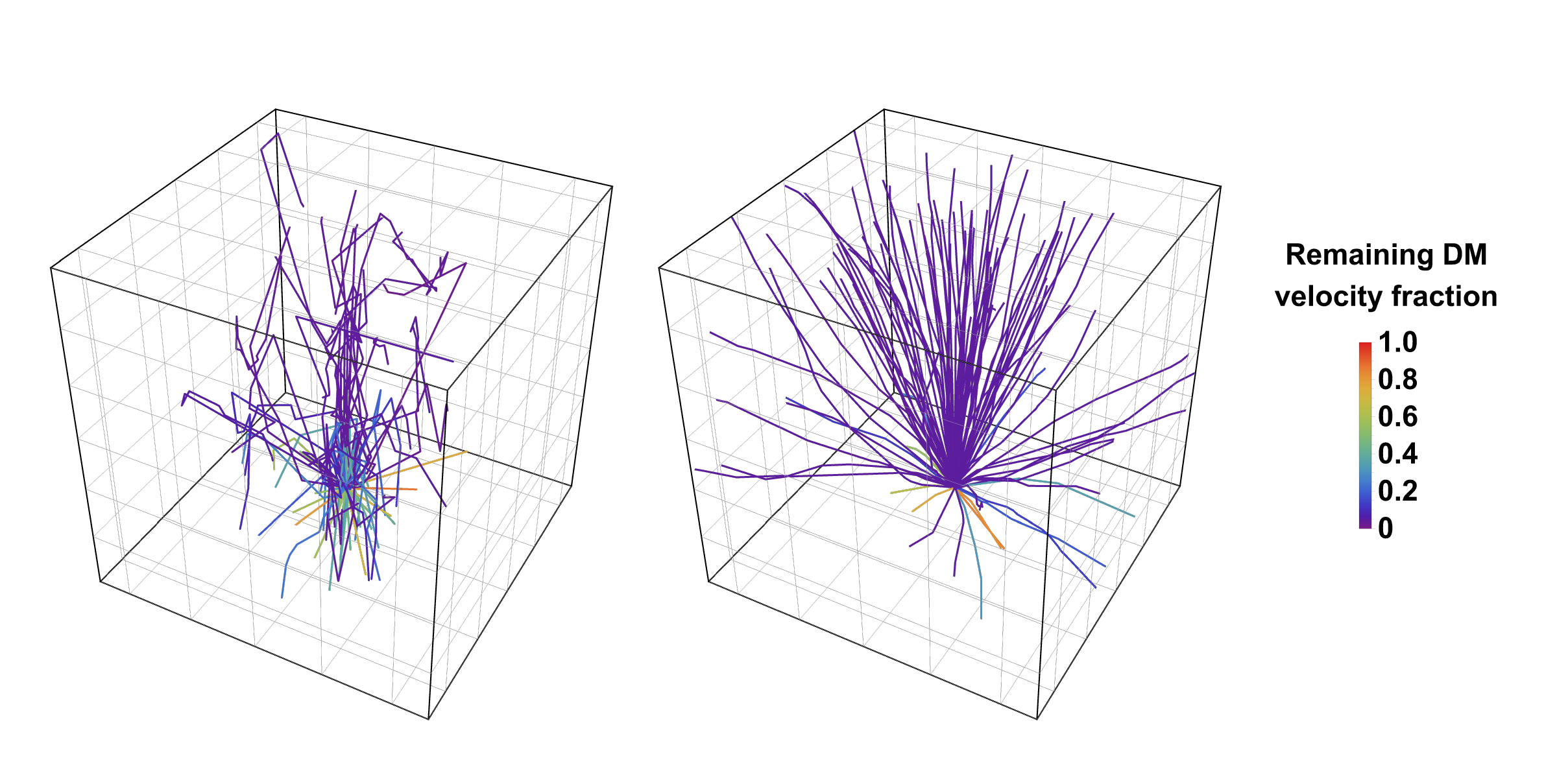}
        \caption{Simulations for DM in Jupiter in the multi-scatter regime; qualitatively similar results are obtained for the Earth. The color of the lines indicates the DM velocity at the end of the run, as a fraction of the incoming DM velocity. At the end of the simulation DM particles have either dropped below the escape velocity to become captured, or have escaped from the object. \textbf{Left:} The diffusive regime where the DM mass is half that of the SM targets. \textbf{Right:} Beginning the departure from the diffuse regime, with DM five times heavier than the SM target mass.}
    \label{fig:simgraphic}
\end{figure*}

\subsubsection{Simulation Results}

Figure~\ref{fig:simgraphic} shows two examples of our simulated outcomes, for DM in Jupiter. In the left panel, the DM is lighter than the SM target, and we clearly see diffusive behavior. That is, upon entering the celestial object, the DM wanders around and does not exhibit dominantly forward motion. In the right plot, we simulated DM that is slightly heavier than the DM target. In this case, the DM more often travels forwards in a straight line, though as the masses are not too different, it is still sometimes reflected, albeit not as often as the lighter DM scenario. As the escape velocity of Jupiter corresponds to about 22 percent of the average incoming DM halo velocity, any velocities below the 0.22 remaining DM velocity fraction (dark blue/violet trajectories) are captured in both panels.

To make our simulations easily applicable to arbitrary celestial objects, we fit analytic functions to the simulation results. The functional form of the fit functions is chosen in order to describe the simulation data with a minimal number of parameters, and the coefficients are determined in a common best-fit minimization across six different celestial objects, ranging in escape velocities from that of the Earth to the Sun. We find therefore that the reflection coefficient can be generalized to 

\begin{align}
f_{\rm cap}(\mu)=
\begin{cases}
f_{\rm cap}^{\rm light}(\mu) &(\mu < \mu_T)\\ \\
\dfrac{ \mu_M f_{\rm cap}^{\rm light}(\mu_T) - \mu_T f_M + \mu \left(f_M - f_{\rm cap}^{\rm light}(\mu_T) \right) }{\mu_M - \mu_T}  &(\mu_T < \mu < \mu_M)\\ \\
\dfrac{\mu}{(\mu - \mu_M) + \mu_M/f_M} &(\mu > \mu_M)
\end{cases}
\label{eq:fcapfull}
\end{align}
for the DM mass regimes as indicated. Here the DM-SM mass ratio $\mu_T$ is determined by $N_{\rm req}(\mu_T) = N_T$, where the critical number of scatterings needed for capture $N_T$ is determined by a fit to the simulations and is given by
\begin{equation}
    N_T \approx 12 + 1.8 \log \left(\sqrt{1 + \frac{v_\chi^2}{v_{\rm esc}^2}}\right).
\end{equation}
This implies a value for $\mu_T$, given by
\begin{equation}
    \mu_T= \frac{\sqrt{2
   \left(v_{\rm esc}^2/(v_\chi^2+v_{\rm esc}^2)\right)^{\frac{1}{N_T}}-1}-\left(v_{\rm esc}^2/(v_\chi^2+v_{\rm esc}^2)\right)^
   {\frac{1}{N_T}}}{\left(v_{\rm esc}^2/(v_\chi^2+v_{\rm esc}^2)\right)^{\frac{1}{N_T}}-1}
\end{equation}
The mass ratio at the second regime change is found to be
\begin{equation}
    \mu_M = 1.56 \, \left[1 -  \frac{1}{1 + 0.52 \log \left(\sqrt{1 + v_\chi^2/v_{\rm esc}^2} \right)} \right],
\end{equation}
and the capture fraction at the second regime change is given by
\begin{equation}
  f_M = 0.22 \, \left[1 +  \frac{3.58}{1 + 0.23 \log \left(\sqrt{1 + v_\chi^2/v_{\rm esc}^2} \right) }\right].
\end{equation}
Our treatment is applicable across a wide range of objects with different escape velocities, up to $v_{\rm esc} \sim 1000 \text{ km/s}$. Objects with larger escape velocities, such as white dwarfs and neutron stars, can be approximated with the formalism we present here, however their high densities can require a more detailed treatment of nuclear effects to obtain more accurate results, and so are beyond the scope of our treatment.

\begin{figure}
    \centering
     \includegraphics[width=0.42\columnwidth]{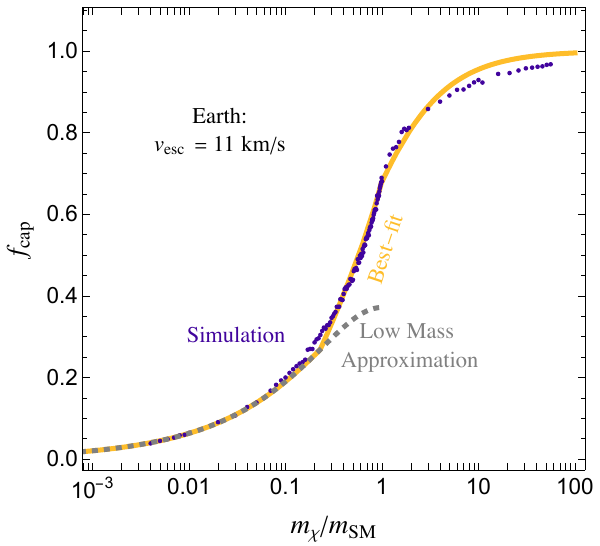}\hspace{5mm}
     \includegraphics[width=0.42\columnwidth]{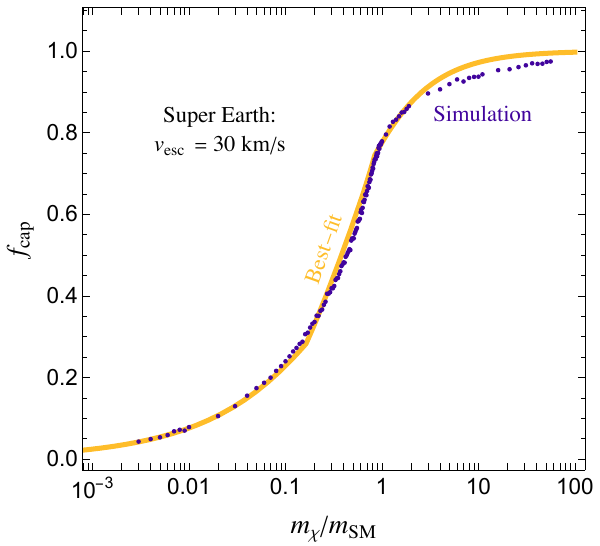}\
          \includegraphics[width=0.42\columnwidth]{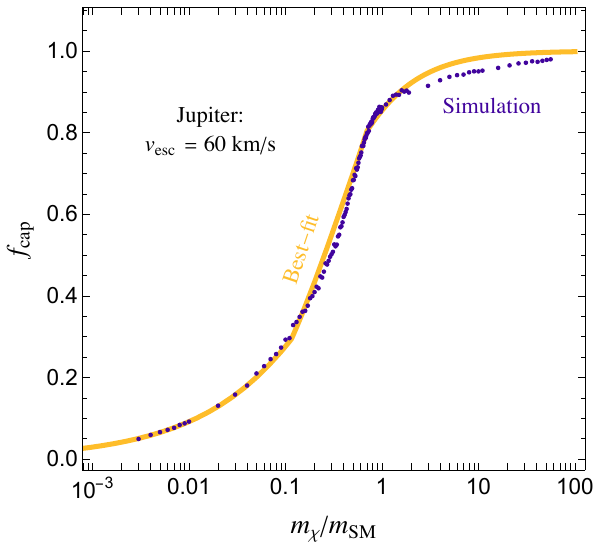}\hspace{5mm}
     \includegraphics[width=0.42\columnwidth]{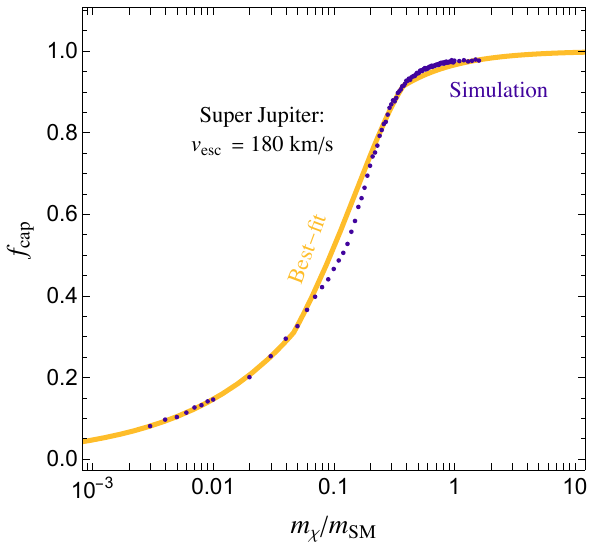}
      \includegraphics[width=0.42\columnwidth]{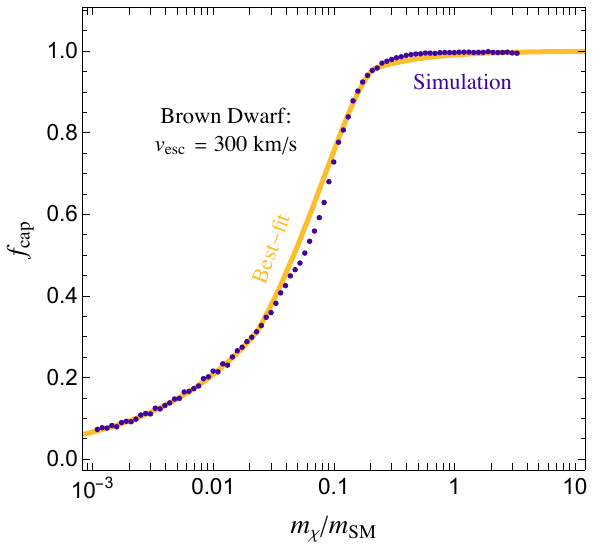}\hspace{5mm}
     \includegraphics[width=0.42\columnwidth]{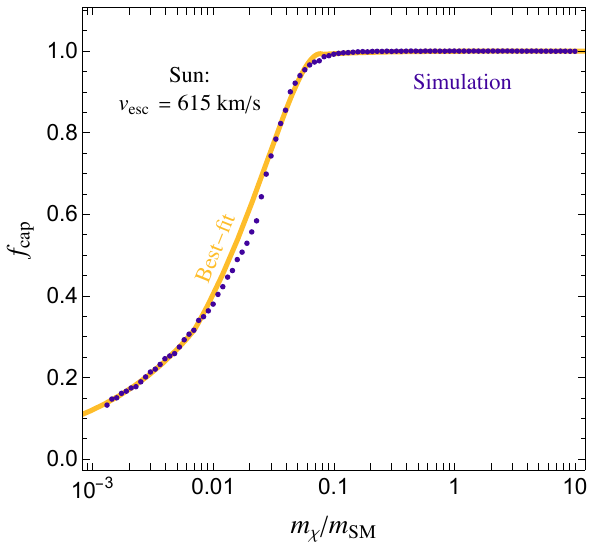}
    \caption{Captured DM fraction taking into account reflection for the Earth, a Super Earth, Jupiter, a Super Jupiter, a Brown Dwarf, and the Sun, with escape velocities as labeled. Our simulations are shown as dots labeled ``Simulation", and our analytic model fits to our simulations are shown as solid lines labeled ``Best-fit". The previous approximation for the low mass limit for the Earth is shown as a gray dashed line, labeled ``Low Mass Approximation", which we match well in the limit of its validity.}
     \label{fig:simreflec}
\end{figure}

Note that the reflection coefficients we approximate here are only relevant for $\tau \gtrsim N_{\rm req}$. At smaller optical depth, the capture is dominated by the first few scatters, and reflection is not relevant. The reflection coefficient is also only relevant in the diffusive regime, which corresponds to the light DM or comparable to target mass regime. However, in the heavy mass regime where $m_\chi\gg m_{\rm SM}$, $f_{\rm cap}\sim1$ only provided that the DM mass does not become so large that it cannot be kinematically stopped. We discuss the separate modification for strongly-interacting and ultra-heavy DM in Sec.~\ref{sec:ballistic}.

Figure~\ref{fig:simreflec} shows a comparison of our simulated data to our functional fits in Eq.~(\ref{eq:fcapfull}) for the reflection coefficient. These are shown for six benchmark celestial objects. The only assumption about the celestial object needed for these plots is the escape velocity of the object as labeled. In the top-left panel, we also compare our results to the analytic expression in Eq.~(\ref{eq:fcap}) as pointed out for the Earth in Ref.~\cite{Neufeld:2018slx} in the light DM limit, shown as ``Low Mass Approximation". Comparing to our simulations, we find that the analytic approximation in Eq.~(\ref{eq:fcap}) is accurate in the limit that the DM is much lighter than the SM target as expected, and corresponds to a critical number of scatterings of $\mathcal{O}(10)$ being required. For DM mass closer to the SM target mass, our analytic fits to our simulations are better approximations to quantify reflection. We also show that high-escape velocity objects such as the Sun have DM capture efficiency loss due to reflection, which was not previously considered.

\subsection{Connecting Reflection Coefficients to Light Dark Matter Capture Rates}

The maximum capture rate in any scenario is given by
\begin{equation}
    C_{\rm max} = C_{\rm geo} \, f_{\rm cap},
    \label{eq:cstrong}
\end{equation}
where $f_{\rm cap}$ is the reflection factor. However, even at very large scattering cross sections, $C_{\rm max}$ will not necessarily be reached due to scattering kinematics. Therefore, the total light DM capture rate is given as a minimum of 
\begin{equation}
    C_{\rm tot}^{\rm light}=\textrm{Min}\left(C^{\rm light} \,, \,C_{\rm max}\right).
    \label{eq:lightmin}
\end{equation}
To evaluate $C^{\rm light}$, we use different numerical tricks depending the cross section size. For $\tau > 3/2$ (the minimum condition for multi-scatter), and $\tau < 100$, we simply use Eq.~(\ref{eq:cnbram}) for  $C^{\rm light}$. To accelerate the computation in the case of very large cross sections above $\tau \approx 100$, we use the fact that the number of scatterings is sufficiently large in order for the sum in Eq~(\ref{eq:CNtotal}) to be replaced by an integral over $N$. Given the simple form of the expression for $p_N(\tau)$ in Eq. (\ref{eq:pNstep}), which is justified at such large optical depth, the integral evaluation is very efficient. Finally, at even larger optical depth above $\tau \approx 10^6$, the evaluation of the integrand leads to a numerical instability. This can be avoided, however, by performing a Taylor expansion of the integrand. In the case of light DM an expansion in $\mu \ll 1$ to order $\mathcal{O}(\mu^4)$ is used,
\begin{align}
& \frac{C_N^{\rm light}}{C_{\rm geo} p_N (\tau)} \approx \min \Biggl\{ 
1, \Biggl|\, 3 \mu N v_{\text{esc}}^4 \Biggl(24 + \mu^3 N^3 \Bigl(64 - 3 \Bigl(9 \frac{v_{\text{esc}}^4}{v_{\chi}^4} \Bigl(\frac{v_{\text{esc}}^2}{v_{\chi}^2} - 6\Bigr) + 76 \frac{v_{\text{esc}}^2}{v_{\chi}^2}\Bigr)\Bigr) \nonumber\\
&-4 \mu^2 \Bigl(3 \mu -1 \Bigr) N^2 \Bigl(\frac{9 v_{\text{esc}}^4}{v_{\chi}^4}-\frac{30 v_{\text{esc}}^2}{v_{\chi}^2}+16 \Bigr) -4 \mu \Bigl(\mu \Bigl(5 \mu -6 \Bigr)+3 \Bigr) N \Bigl(\frac{3 v_{\text{esc}}^2}{v_{\chi}^2}-4\Bigr) \nonumber\\
&+ 8 \mu \Bigl(\mu -3 \Bigr)\Biggr) \times \Biggl(8 \Bigl(3 v_{\text{esc}}^2 v_{\chi}^2 + 2 v_{\chi}^4\Bigr)\Biggr)^{-1}\Biggr| \Biggr\}. && 
\label{eq:cnlight}
\end{align}
This procedure results in a sub-percent accuracy at a substantial increase in calculation speed.

\section{Discussion of the Intermediate Interaction Regime}
\label{sec:transition} 

\begin{figure}[H]
\centering
\includegraphics[width=0.5\columnwidth]{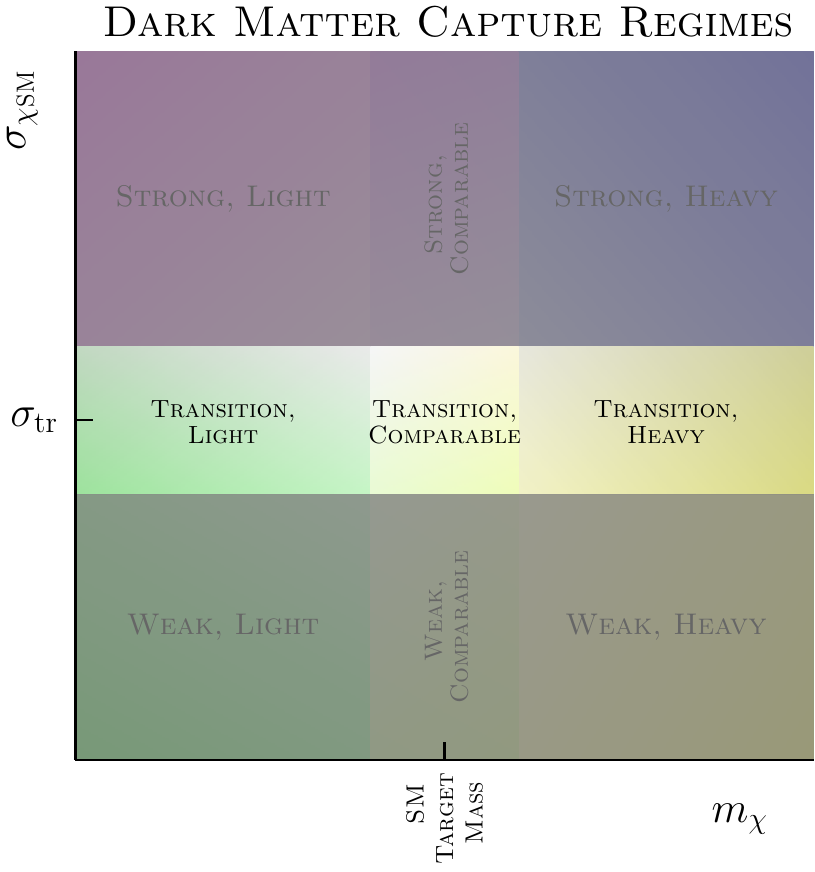}
    \caption{The transition between weak and strong interaction regimes, as discussed in this section.}
    \label{fig:capturegridlighttrans}
\end{figure}

We now consider the intermediate interaction regime, which we mark by the transition cross section $\sigma_{\rm tr}$, which has been previously labeled $\sigma_{\rm sat}$ in some works in the literature. We first discuss the meaning of this cross section and advocate for the term of a ``transition cross section" rather than the previous term ``saturation cross section", then discuss the capture behavior when the cross sections switch from weak to strong interactions.

\subsection{The Meaning of the ``Saturation'' Cross Section}
\label{sec:sat}

 We emphasize that the physical meaning of  $\sigma_{\rm tr}$ (previously sometimes called $\sigma_{\rm sat}$ in the literature), where the optical depth is $\tau =3/2$, is simply that the mean free path becomes of order the size of the object, and therefore larger cross sections than $\sigma_{\rm tr}$ enter the multiscatter regime, and smaller cross sections than $\sigma_{\rm tr}$ correspond to the single scatter regime. $\sigma_{\rm tr}$ should not be interpreted as the cross section generically resulting in the maximum capture rate, as we now discuss.

There is some ambiguity in the literature about the meaning of the saturation cross section, $\sigma_{\rm sat}$ (which we are now calling $\sigma_{\rm tr}$). This quantity was first referred to as a saturation cross section in Ref.~\cite{Bramante:2017xlb}, which studied DM capture in neutron stars. Because neutron stars have extreme escape velocities, they are very efficient at capturing DM. Provided the DM mass is not too different compared to the SM target mass this means that the maximum capture rate is usually achieved with a single scatter, such that the maximal capture and the ``saturation'' of the cross section is reached right before the multi-scatter regime kicks in. This is because neutron stars' capture efficiency implies adding any scatters more than roughly once will not help them capture DM any further. Therefore, calling $\sigma_{\rm tr}$ a saturation cross section is valid when applied to its original use in Ref.~\cite{Bramante:2017xlb} for neutron stars, although it is not the physical meaning of the cross section. In fact, for any object where the escape velocity is higher than the DM halo velocity, $\sigma_{\rm tr}$ can correspond to the maximum capture rate provided the DM mass is not too far from the SM target mass.

However, we emphasize that this is not true in objects with lower escape velocities compared to the DM velocity, such as Jupiter or the Earth. For objects with escape velocities below the DM halo velocity, because they are not very efficient at capturing DM, usually many scatters are required to slow down and capture the DM, and so larger cross sections than $\sigma_{\rm tr}$ are often required to capture the bulk of the DM.

\begin{figure}[t!]
\centering
\includegraphics[width=0.6\columnwidth]{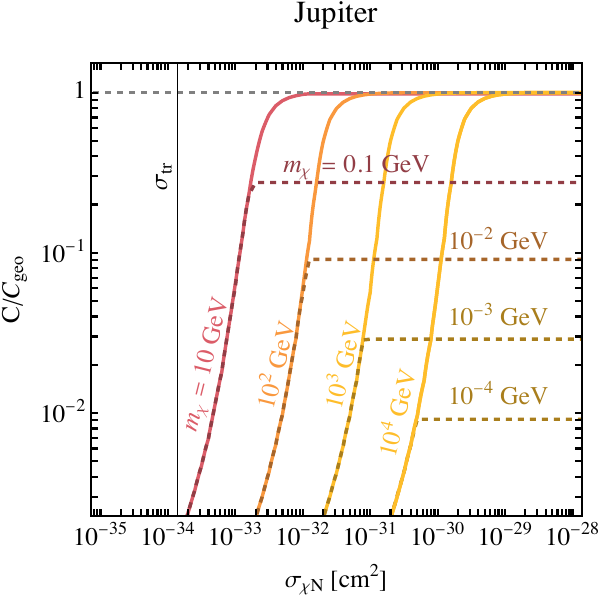}
    \caption{Fraction of DM particles passing through Jupiter that are captured as functions of the DM-SM scattering cross section, for several DM masses as labeled. The transition cross section $\sigma_{\rm tr}$ for Jupiter is shown as a vertical line, and the gray dashed line at $C/C_{\rm geo}=1$ corresponds to all the DM passing through Jupiter being captured.}
    \label{fig:transition}
\end{figure}

Figure~\ref{fig:transition} makes this point explicit, and shows the fraction of DM particles passing through Jupiter that are captured as functions of the DM-SM scattering cross section. Here Jupiter is chosen as an example of a low-escape velocity object to illustrate the meaning of $\sigma_{\rm tr}$; qualitatively similar results are obtained for other low escape velocity objects such as the Earth. Fig.~\ref{fig:transition} shows a range of DM masses: the solid lines are for DM heavier than target, and the dashed lines are the inverse of the solid lines, making them lighter than the SM target and symmetrically far away from the target mass. As the DM scattering is most kinematically efficient when the DM is comparable to the SM mass, taking DM masses that are symmetrically lighter or heavier than the target requires multiscattering for capture. The symmetry in the behaviour of DM mass both the heavier and lighter than the target show the identical scaling of the capture rate, apart from the fact that at very high cross sections light DM is reflected, which explains the departure of the dashed curves for the light DM at high cross sections.

In Fig.~\ref{fig:transition}, we see that compared to the transition cross section $\sigma_{\rm tr}\approx10^{-34}~$cm$^2$ for Jupiter, the maximum capture rate is actually achieved at cross sections higher than the transition cross section, simply because Jupiter is not very efficient at capturing DM due to its escape velocity being lower than the local DM halo velocity, and therefore requires the multiscatter regime in order to capture the bulk of the DM. This clearly illustrates that the ``saturation cross section" is not the physical meaning of $\sigma_{\rm tr} = {\pi R^2}/{N_n}$, and taking the ``saturation cross section" as the cross section providing the maximum DM capture rate, would yield capture rates orders of magnitude too small.

Given the broadened interested in DM in a variety of celestial objects, and the true physical meaning of this cross section, we advocate for calling this cross section instead the transition cross section, which in this work we are calling $\sigma_{\rm tr}$. As we described above, $\sigma_{\rm tr} = {\pi R^2}/{N_n}$ marks the transition between the single and multiscatter regimes for DM capture in arbitrary celestial objects, it is not generically the cross section where the maximum capture rate will be found. Indeed, note that already in some works, e.g. Ref.~\cite{Bell:2021fye}, the term ``threshold cross section" and denoted also by $\sigma_{\rm tr}$ was also already used, instead of ``saturation cross section".

\section{Ballistic Dark Matter Regime}
\label{sec:ballistic}

\begin{figure}[h!]
\centering
\includegraphics[width=0.5\columnwidth]{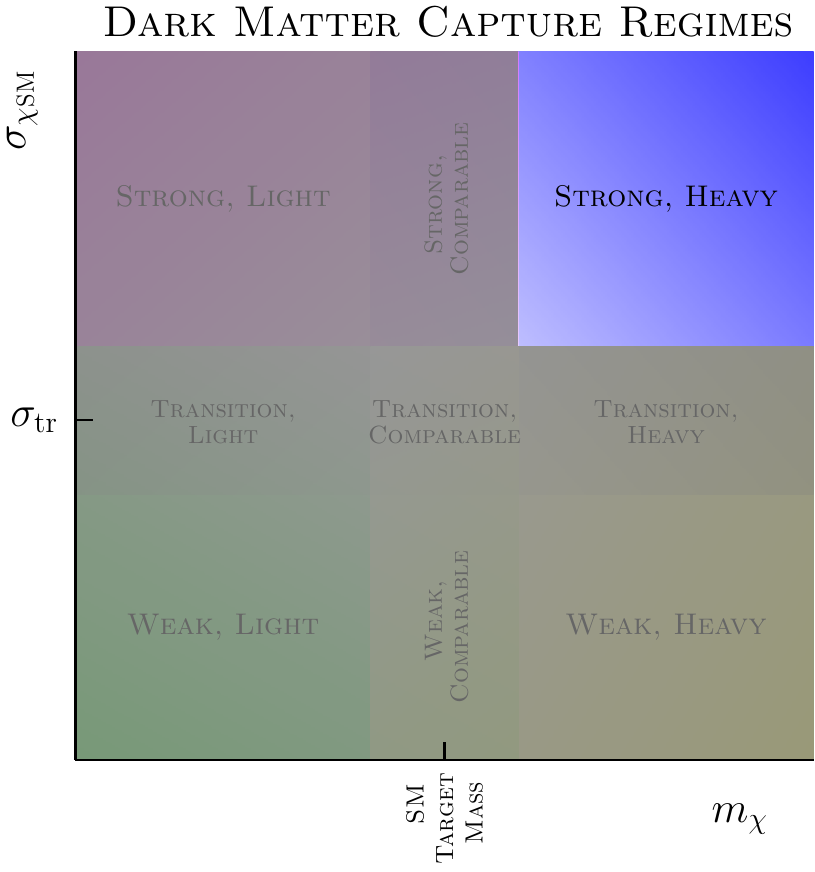}
    \caption{The heavy DM mass regime, with strong interactions discussed in this section. }
    \label{fig:capturegridheavystrong}
\end{figure}

As discussed above, when DM is heavy, the straight forward-line motion is accurate for the DM trajectory inside a celestial object. However, if DM becomes substantially heavier than the SM target mass, it can have another problem. This is simply that it can have so much momentum that the celestial-body has insufficient stopping power, and the ultra-heavy DM blasts out the other end of the celestial body without being captured, even in the strong interaction regime. This can occur if the optical depth of DM in a given object does not allow the required number of scatters for capture. In Sec.~\ref{sec:sum}, we included an improved condition, similar to the one introduced in Ref.~\cite{Beacom:2006tt} for the Earth, which covers this scenario.

In addition, when the DM mass is much heavier than the target mass, and the scattering is fully forward dominated, there is a finite number of targets that the particle can encounter on its way through the object. This number is given by $N_c \sim  N_n^{1/3}$, with $N_n$ being the total target number in the object. Thus, we consider only physically relevant values of the optical depth $\tau_{\rm physical} = \min \left[\tau ,  N_c \right]$ in this regime.

Finally, to calculate DM capture in the case of heavy DM and very large cross sections, we use similar approximations to calculating $C_N^{\rm light}$ in the previous section. For optical depths above $\tau \approx 100$, we replace the sum in Eq.~(\ref{eq:CNtotal}) by the integral, and use the $p_N(\tau)$ from Eq. (\ref{eq:pNstep}). Again at optical depth above $\tau \approx 10^6$, the evaluation of the integrand leads to a numerical instability, and is avoided by performing a Taylor expansion of the integrand. For heavy DM with $\mu \gg 1$ we go to order $\mathcal{O}(\mu^4)$, which leads to
\begin{align}
\frac{C_N^{\rm heavy}}{ C_{\rm geo} p_N (\tau)} & \approx \min \Biggl\{ 
1 , \Biggl|\,  3 N v_{\text{esc}}^4 \Biggl(8 \mu (3 (\mu -1) \mu +1) + N^3 \Biggl(64- 3 \Biggl(9 \frac{v_{\text{esc}}^4}{v_{\chi }^4} \Biggl(\frac{v_{\text{esc}}^2}{v_{\chi}^2}-6\Biggr)+76 \frac{v_{\text{esc}}^2}{v_{\chi }^2}\Biggr)\Biggr) \nonumber \\
&+4 (\mu -3) N^2 \Biggl(\frac{9 v_{\text{esc}}^4}{v_{\chi }^4}-\frac{30 v_{\text{esc}}^2}{v_{\chi }^2}+16\Biggr) \nonumber \\
&-4 (3 (\mu -2) \mu +5) N \Biggl(\frac{3 v_{\text{esc}}^2}{v_{\chi }^2}-4\Biggr) \Biggr) \times \Biggl(8 \mu ^4 \Biggl(3 v_{\text{esc}}^2 v_{\chi }^2+2 v_{\chi }^4\Biggr)\Biggr)^{-1}\Biggr| \Biggr\}. &&
\end{align}
These simplifications for the heavy DM regime result in a sub-percent accuracy with a substantial increase in calculation speed.

\section{Capture for Benchmark Objects}
\label{sec:objects}

\begin{figure*}[!t]
    \centering
     \includegraphics[width=0.48\columnwidth]{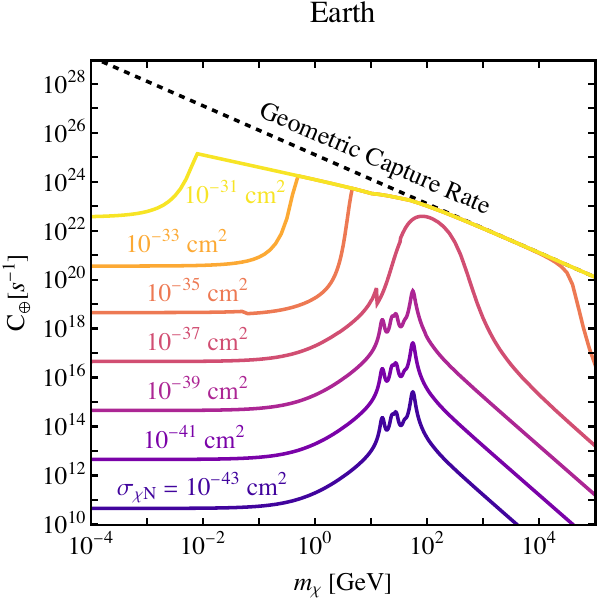}\hspace{5mm}
     \includegraphics[width=0.48\columnwidth]{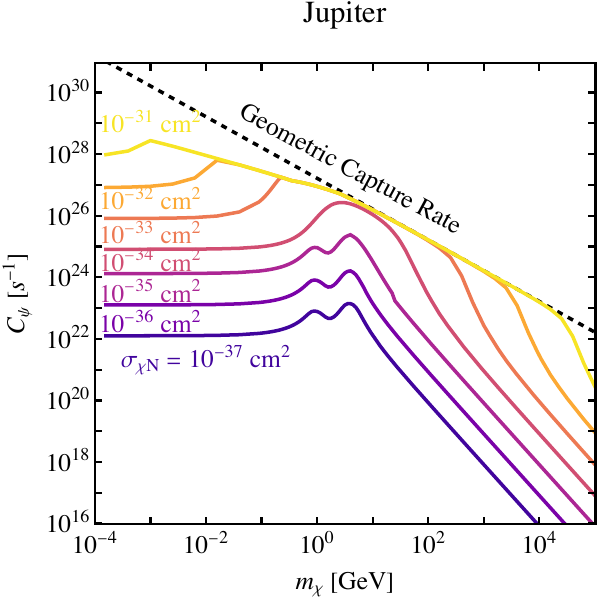}\vspace{7mm}
          \includegraphics[width=0.48\columnwidth]{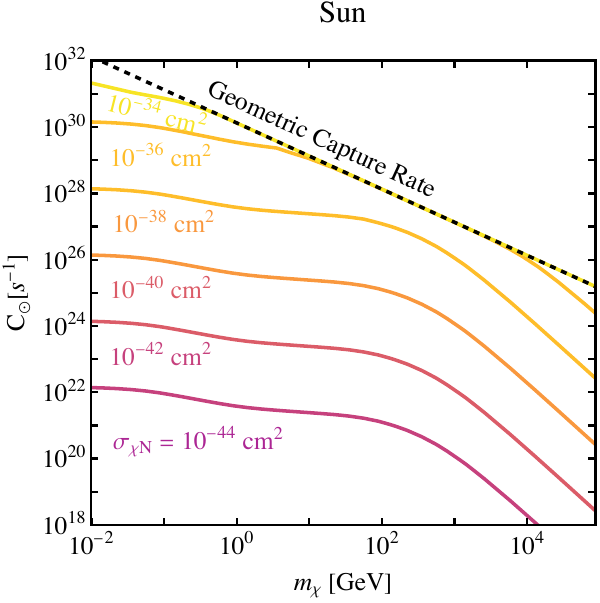}\hspace{5mm}
     \includegraphics[width=0.48\columnwidth]{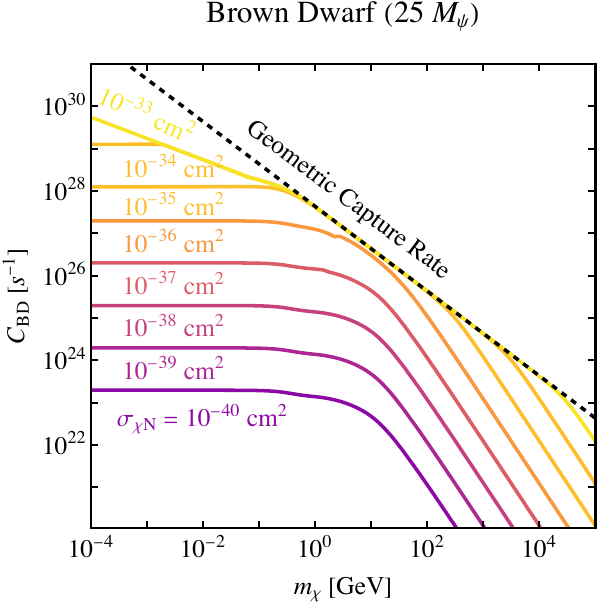}
    \caption{DM capture rates as a function of DM mass in the Earth, Jupiter, Sun, and a benchmark Brown Dwarf, for a range of DM-nucleon scattering cross sections $\sigma_{\chi N}$ as labelled. The dashed line corresponds to the geometric capture rate, which corresponds to all the DM passing through the object being captured.}
    \label{fig:objects}
\end{figure*}

\begin{figure*}[!t]
    \centering
     \includegraphics[width=0.48\columnwidth]{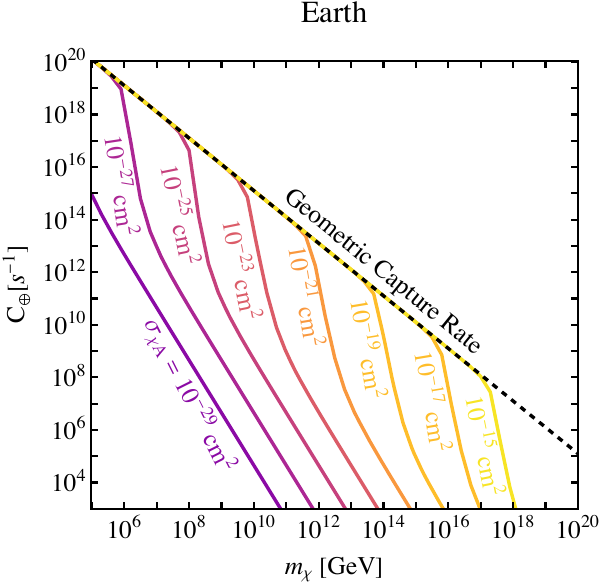}\hspace{5mm}
     \includegraphics[width=0.48\columnwidth]{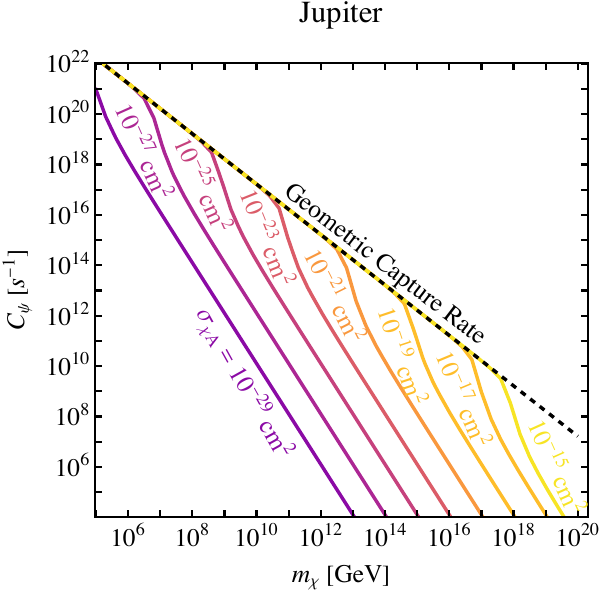}\vspace{7mm}
          \includegraphics[width=0.48\columnwidth]{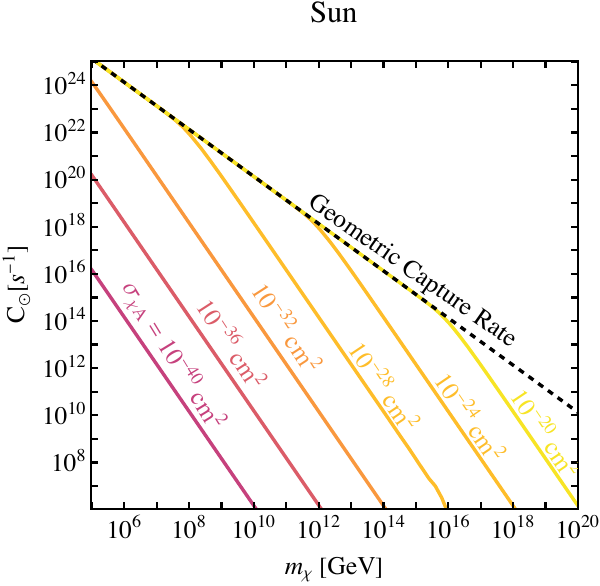}\hspace{5mm}
     \includegraphics[width=0.48\columnwidth]{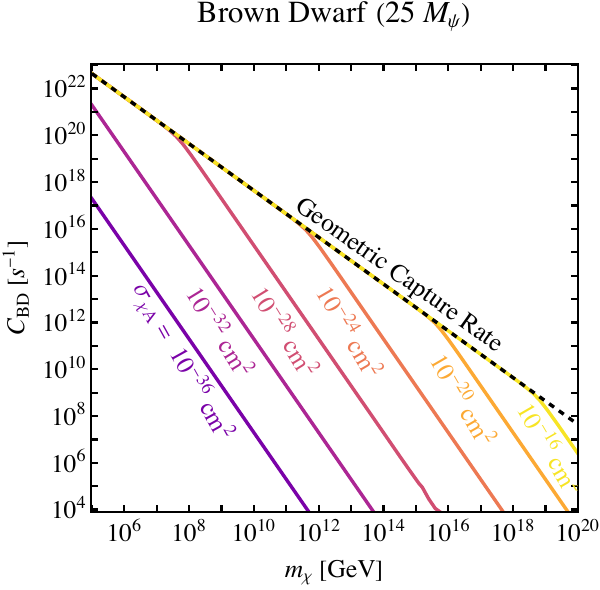}
    \caption{DM capture rates as a function of DM mass in the Earth, Jupiter, Sun, and a benchmark Brown Dwarf, for a range of DM-nucleus scattering cross sections $\sigma_{\chi A}$ as labelled. Note that we show the total scattering cross sections of DM with the nuclei $\sigma_{\chi A}$ here. Depending on the model a connection to the nucleon cross sections can be made, however at large cross section values this might not be possible due to breakdown of the Born approximation. The dashed line corresponds to the geometric capture rate, which corresponds to all the DM passing through the object being captured. Note that we have not included an atomic mass enhancement, see text for discussion.}
    \label{fig:objectsheavy}
\end{figure*}

Figure~\ref{fig:objects} shows our DM capture rates as a function of DM mass for fixed cross sections, for four benchmark objects: the Earth, the Sun, Jupiter, and a Brown Dwarf with 25 Jupiter masses and the radius of Jupiter. Here these objects are assumed to be in the local position with $\rho_\chi=0.4$~GeV/cm$^3$ and $v_\chi=270$~km/s. We assume $75\%$ H and  $25\%$ He for both Jupiter and the brown dwarf. For the Earth we use the six most abundant elements ($32\%$ Fe, $29\%$ O, $15\%$ Mg, $14\%$ Si, $1.7\%$ Ca, $1.5\%$ Al) from Ref.~\cite{Bramante:2019fhi}. These numbers are determined by taking a mantle mass fraction of $67\%$ with the rest of the mass being the Earth core. For the Sun we take the volume averaged abundances of ($68.6\%$ H, $29.9\%$ He, $0.64\%$ O, $0.19\%$ C, $0.15\%$ Ne, $0.13 \%$ Fe) from Ref.~\cite{Vinyoles:2016djt}. 

The cross sections shown in Fig.~\ref{fig:objects} are DM-nucleon, such that appropriate mass number $A$ scalings are taken into account. These rates include all the effects we have considered: the DM motion direction be it diffusive or ballistic, and a range of DM mass and cross section regimes. For objects with escape velocities less than the DM halo velocity, namely the Earth and Jupiter, we observe that the geometric capture rate is not obtained unless the DM is comparable in mass to the SM target mass. For light DM, the capture rates decrease from the maximum value with the scaling we found from our simulations. For the high-escape velocity objects, the Sun and a local Brown Dwarf, lighter DM can reach the geometric capture rate compared to low-escape velocity objects. Eventually, for cross sections much less than the cross section corresponding to the geometric capture rate, for all objects the capture rate eventually goes flat. This happens in the optically thin regime as the mass capture rate for DM lighter than the target scales as $m_\chi$, while the number density scales as $1/m_\chi$. Thus the contributions cancel leading to a mass independent DM number capture rate. However for the heavy DM mass case, the mass capture rate scales as 1/$m_\chi$, such that turning this into a DM number capture rate leads to a 1/$m_\chi^2$ scaling. Both heavy and light cases are equally inefficient at capture away from the target mass, which can be verified by plotting instead the mass capture rate, which does not take into account the relative boost or decrease in DM number densities for light or heavy DM masses respectively.

Compared to the capture rates for the Earth in Ref.~\cite{Bramante:2022pmn}, we find very different results away from the SM target mass. Ref.~\cite{Bramante:2022pmn} states that their simulations were not sufficiently populated once the capture rate was five orders of magnitude or more below the geometric capture rate at a given DM mass. Ref.~\cite{Bramante:2022pmn} therefore extrapolated their simulation results for larger capture rates to smaller capture rates, fitting the downwards slope as a line that continued downwards indefinitely. However, in the analytical treatment we find this downwards slope does not continue indefinitely, and instead does turn around and become flat (independent of DM mass) for light DM at a fixed cross section. The difference between our analytical results is due to our inclusion of the low-velocity tail of the DM distribution incoming at Earth, while Ref.~\cite{Bramante:2022pmn} treats all DM particles as being sped up to the solar escape velocity at Earth's position, which effectively removes the DM velocity distribution below $\sim42$~km/s at Earth's position. However, it was shown previously that the combined effects of gravitational diffusion from solar system bodies such as the Sun, Jupiter, Mercury, and Venus, as well as elastic scattering from these bodies, leads to about the same DM phase-space distribution as the scenario where the celestial-body is not placed deep in the gravitational well of the solar system; see Ref.~\cite{Sivertsson:2012qj} as well as the earlier Refs.~\cite{1991ApJ...368..610G, Peter:2009mm, Peter:2009mk,Peter:2009mi}. Therefore, assuming solar system objects are free objects is an appropriate approximation for the purposes of DM capture, assuming that gravitational equilibrium has been reached. This means that the low-velocity tail of the DM distribution is expected to be well populated~\cite{Sivertsson:2012qj,1991ApJ...368..610G, Peter:2009mm, Peter:2009mk, Peter:2009mi}, and the single-scatter regime is therefore not expected to be suppressed.

Figure~\ref{fig:objectsheavy} shows capture rates for the same objects in Fig.~\ref{fig:objects}, but for larger DM masses. The cross sections are quoted as for DM-nucleus scattering, though importantly note that we have not included any atomic mass number enhancements in this figure. This is because these cross sections are so large that they are in the regime that the Born approximation breaks down, see for example the discussion in Refs.~\cite{Digman:2019wdm,Xu:2020qjk}. As large DM masses are very far away from the SM target masses, capture is very inefficient, and many scatters are needed to slow the DM down sufficiently. Heavy DM can readily just blast out the other side of the celestial object if there is not sufficient stopping power, so much larger cross sections are required to retain the geometric flux of DM particles impinging on the object. Note that not including the limited stopping power would give too large capture rates (as per e.g. Ref.~\cite{Ray:2023auh}, which assumes geometric capture even for $\tau>N_{\rm req}$). We have also included the limiting case where the number of targets in the system is insufficient for the required optical depth, which is most evident in the $10^{-15}$~cm$^2$ line for Jupiter. It takes such extreme cross sections for this condition to be relevant, which is why the other cross section lines in the Jupiter plot do not have the same shape.

Note that in Fig.~\ref{fig:objects} and Fig.~\ref{fig:objectsheavy}, we do not show objects such as neutron stars and white dwarfs. This is simply because they require extra ingredients due to their very high densities, as we already noted earlier, though our setup can be used at an approximate level for these objects, and equate to zero-temperature capture rates with neglected nuclear and degeneracy effects. Given the thorough study of DM capture in neutron stars and white dwarfs already existing in the literature~\cite{Bramante:2017xlb,
Bell:2019pyc,Bell:2020jou,Bell:2020obw,Bell:2021fye}, we do not investigate these objects further. For all objects we have in fact not included any thermal motion of the targets, at very light DM masses these can become relevant, especially for the Sun~\cite{Garani:2017jcj}.

In Fig.~\ref{fig:objects} and Fig.~\ref{fig:objectsheavy}, we assume that there are no appreciable long-range attractive DM-SM interactions. This is important to note as the presence of these can increase the predicted escape velocities of celestial objects, which can lead to objects which were previously considered to be ``low-escape velocity" objects, to not be at all, such that they reach their geometric capture rate closer to the transition cross section. The importance of this effect has recently been discussed in the context of DM evaporation, where it was shown that the DM evaporation mass is highly-model dependent~\cite{Acevedo:2023owd}. In a similar vein, the capture rates can be highly-model dependent, however the formalism to calculate the capture rates remains largely the same, with the celestial-body escape velocity just updated (provided that the escape velocity does not become relativistic).

In our treatment the density profiles of the celestial objects are not taken into account, as this is expected to have a subdominant effect on the capture rate, and is beyond the scope of our current study. In the same matter the DM evaporation mass can not be calculated with our package, as it depends detailed properties of the celestial object, such as density and temperature profile, and is in general more model dependent~\cite{Acevedo:2023owd}.

\section{Summary and Conclusions}
\label{sec:conclusion}

DM can be captured by a range of celestial objects, providing observable signals. Calculating signal detectability first requires an accurate calculation of the DM capture rate. Previous works have largely focused on capture regimes where DM is heavier than the SM celestial-body target, and in dense objects such as neutron stars. Recently however, interest has grown in DM lighter or comparable to the SM target mass, as this parameter space still remains relatively unexplored by direct detection experiments, and can produce highly detectable signals in celestial objects in a range of objects including those which require different treatment to ultra-dense neutron stars. We have discussed a range of effects relevant for DM capture across a wide range of celestial objects, focusing on DM-nucleon interactions, and building upon and improving frameworks in the literature. As an important part of this work, we have released our multi-regime calculations as a public package available in both Python and Mathematica versions, called \texttt{Asteria}~\cite{asteria}.

In the strong interaction regime, we performed new simulations that complement existing works. We investigated the behaviour of light and comparable mass DM in the diffusive regime, to include changes in the DM trajectory compared to the previous standard forwards linear motion scenario, determining the fraction of DM particles expected to be captured or ejected during scattering. We fit our simulated capture rates with analytic functions that can be applied to many objects, such as the Earth, Jupiter, a Brown Dwarf, or the Sun.

We discussed some improvements to the capture summation treatment, which allow for easier and faster numerical treatment in the large cross section regime. We also introduced a condition for the number of scatters for truncation of the capture summation, which is relevant for all strongly interacting regimes, but particularly for the ultra-heavy DM scenario where the celestial-body stopping power limits DM capture. Furthermore, our treatment of the DM reflection in the light DM regime provides a theoretical upper bound on the DM capture rate based purely on the kinematics of the energy transfer, a finding that will affect a number of light DM searches.

Our capture rate treatment and package covers light DM with diffusive motion which can random walk back out of celestial objects, light DM with weak interactions, heavy DM which might blast out the end of celestial objects due to lacking stopping power even for very large interaction cross sections, limited number of targets available for capture residing in celestial objects, and how to treat intermediate regimes in between. Consequently our treatment and numerical package can be applied to a wide range of active and upcoming searches for DM in celestial bodies, providing the fundamental quantity required: the capture rate.

\section*{Acknowledgments} 

We thank ChatGPT for coming up with our package name. ChatGPT's rationale was, verbatim: \textit{``Asteria is a good name for your package because it's a Greek goddess associated with the stars and the night sky. The name Asteria comes from the Greek word ``aster," meaning star, and so it is fitting for a package that deals with the capture of dark matter in celestial objects. As the goddess of the night sky, Asteria is also associated with the mysteries of the universe and the unknown, which is appropriate for a package that deals with the elusive nature of dark matter."} We also thank humans for their helpful discussions and comments, specifically Javier Acevedo, Joe Bramante, Chris Cappiello, Djuna Croon, and Annika Peter. RKL is supported in part by the U.S. Department of Energy under Contract DE-AC02-76SF00515. 

\appendix

\providecommand{\href}[2]{#2}\begingroup\raggedright\endgroup

\end{document}